\documentclass[aps,floats,twocolumn,showpacs]{revtex4}
\usepackage{graphicx,epstopdf}
\begin{document}

\title{Superconducting nanowire quantum interference device based on Nb ultrathin films deposited on self-assembled porous Si templates}
\author{C. Cirillo$^1$, S. L. Prischepa$^{2}$, M. Trezza$^1$, V. P. Bondarenko$^{2}$, and C. Attanasio$^1$}

\affiliation{$^1$CNR-SPIN Salerno and Dipartimento di Fisica \lq\lq E.R. Caianiello\rq\rq, Universit\`{a} degli Studi di Salerno, via Giovanni Paolo II, Fisciano (SA) I-84084, Italy\\
$^2$Belarusian State University of Informatics and Radioelectronics, P. Browka 6, Minsk 220013, Belarus}

\date{\today}

\begin{abstract}

Magnetoresistance oscillations were observed on networks of superconducting ultrathin Nb nanowires presenting evidences of either thermal or quantum activated phase slips. The magnetic transport data, discussed in the framework of different scenarios, reveal that the system behaves coherently in the temperature range where the contribution of the fluctuations is important. 

\end{abstract}


\maketitle

\section{Introduction}

Superconducting materials having nanometric characteristic dimensions represent a huge field of investigation, which became accessible only very recently. Properties of superconducting nanoparticles and nanowires (NWs) \cite{Tinkham,nanopart,AruPhysRep,BezBook} are only two examples of how rich the physics revealed by these systems can be. In particular, in the last years transport properties of superconducting NWs have been intensively investigated both to address fundamental issues \cite{AruPhysRep,BezBook,Rodrigo}, as well as to find new applications in superconducting electronics \cite{Hopkins,Delacour,Murphy,Weides}. Similarly, superconducting nanowires in a Dayem bridge configuration and, in general, multiply-connected nanowire arrays may be the core of superconducting devices working as magnetometers and radiation detectors \cite{Granata,BezryadinScience,Pekker,Hansma}. Moreover recently, the proposed duality between the Josephson junction and the quantum phase slip (QPS) effect \cite{Nazarov} strengthened the interest in these low-dimensional systems, which, under appropriate conditions, could show coherent QPS \cite{Astafiev} paving the way to the realization of QPS qubit \cite{Mooij,Belzig}, QPS transistors \cite{Hongisto}, as well as quantum current standards \cite{Webster}.
The approach to NWs fabrication became in itself a research field, spanning from ion-beam design \cite{AruNanotech,Tettamanzi}, to molecular templating \cite{BezrNanotech}, and deposition inside \cite{Tian} or on top \cite{Luo} of nanoporous self-assembled substrates. The latter method, in particular, results to be specially appealing since it allows the fabrication of patterned nanostructures in a single step, rapidly, cheaply, and with high reproducibility on a large scale area. 

Independently of the particular fabrication method, it was widely demonstrated that when the wire width, $w$, approaches the superconducting coherence length, $\xi$, both thermal (TAPS) \cite{BezAPL,Luo} and quantum \cite{Giordano,BezryadinNat,Lau} activated phase slips play an increasing role on the transport. Moreover, not only the well known broad resistive transitions and hysteretic stepwise current-voltage characteristics are observed \cite{BezAPL,Lau}, but often periodic magnetoresistance (MR) oscillations are reported. This last phenomenon concerns structures of different topology, namely systems with two or more nanowires (the so-called nanowire networks) \cite{BezryadinScience,Pekker,Sochnikov,Baturina,Zhang1,Zhang2}, as well as single nanowires \cite{Herzog,Johansson,Patel1,Wang1,Wang2,Patel2,Lehtinen1,Mills}, or systems having an intermediate geometry, such as ladder structures \cite{Peeters}.

Most of the MR results \cite{Baturina,Luo,Zhang1,Zhang2} displayed by nanowire networks are well interpreted as commensurability and frustration effects \cite{Pannetier,Kato} with the period of the resistance oscillations being an integer or a specific fractional multiple of $\Delta H= \Phi_{0}/S$, where $\Phi_{0}$ is the flux quantum and $S$ is the area enclosed by the single wire loop. This is a direct consequence of the fluxoid quantization, as predicted by Little and Parks (LP) for superconducting cylinders \cite{Little}. Recently, in Nb ladder structures the oscillations were successfully ascribed to the interplay between bias and screening currents \cite{Peeters}. On the other hand, the oscillations observed on singly connected wires were interpreted in the framework of different scenarios, namely screening currents circulating around the samples grains as in the case of granular Sn and NbN wires \cite{Herzog,Patel2}, non-uniform sample thickness \cite{Johansson,Wang1,Wang2}, vortex-row confinement effects \cite{Patel1}, and presence of randomly distributed pinning centers creating channels for the vortex flow \cite{Mills}, but their origin in some cases is still under debates \cite{Lehtinen1}.

This work mainly focuses on magnetoresistence measurements performed on superconducting Nb nanowire networks. The samples, whose resistive transitions are dominated by either thermal or quantum fluctuations, exhibit oscillatory behavior of the resistance as a function of the field, $R(H)$, at temperatures very close to the superconducting critical temperature, $T_c$. These oscillations, which cannot be simply explained as due to the Little-Parks effect, 
were instead interpreted in the framework of a multiple-current-path context \cite{Hansma}. Moreover, a parallel between this system and a device based on two nanowires, which acts as a sensitive superconducting phase gradiometer, \cite{BezryadinScience,Pekker}, was drawn. 

\section{Experimental methods}

As previously described, a convenient way to produce arrays of superconducting NWs is to exploit the regular and nanometric structures of self-assembled templates. Indeed, here Porous Silicon (PS) \cite{Pavesi} was used as substrate for depositing high quality ultrathin Nb films. PS is a highly technological material employed in several fields, being used, for instance, in sensors, light emitting devices \cite{Pavesi}, or lithium batteries \cite{Ge}. It is commonly produced by electrochemical etching of Silicon single crystal in a solution of hydrofluoric acid, under appropriate conditions of current, light and temperature. Producing PS with given required pore dimensions and thickness is a non-trivial process to control. Since the key parameters to monitor are the Silicon doping, the concentration of HF in the electrolyte, and the electrical anodization regime \cite{PSL} investigations for a better tuning of the procedures are continuously active. In the present work n+ type antimony doped 100 mm monocrystalline silicon wafers with (100) orientation and $0.01~\Omega \times$cm resistivity were used as the initial substrates. Silicon wafers were cleaned by using the RCA solution, dried in the centrifuge and cut into a number of rectangular 9 cm$^2$ samples. Just before PS formation the experimental samples were immersed into 5\% HF solution to remove the native oxide. Immediately after that, the Si samples were disposed in a standard electrolytic cell made of Teflon. Uniform PS layers were formed by an electrochemical anodization of silicon samples in 1:3:1 =HF:H$_2$O:(CH$_3$)$_2$CHOH solution \cite{Dolgiy12}. The anodized samples were rinsed in the deionized water for 5 min. A spectrally pure graphite disk was used as a contact electrode to the back side of the samples during the electrochemical treatment. Platinum spiral wire was used as a cathode electrode. Anodization was carried out at a current density $j=10-13$ mA/cm$^2$ for 200 s. This regime provided the formation of uniform PS layers with thickness of 10 $\mu$m and porosity of 50\% \cite{Panarin10}. 

\begin{table}
\begin{tabular}{ccccccc}
sample & $d_{Nb}$(nm) & $\oslash $(nm) & $\Lambda $(nm) & $w$(nm) & $\sigma$(nm) & $T^{10\%}_{c}$(K) \\
\hline
A & 9.0 & 5 & 10 & 5 & 6.7 & 3.00 \\

B & 9.0 & 10 & 40 & 30 & 16.4 & 3.00 \\

C & 8.5 & 10 & 40 & 30 & 16.0 & 3.65 \\
\end{tabular}
\vspace{4mm} \caption{Samples characteristic parameters: $d_{Nb}$ is the Nb thickness, $\oslash$ indicates the pore diameter, $\Lambda$ the interpore spacing, $w$ the wire width, $\sigma$ the wire effective diameter, and $T^{10\%}_{c}$ the critical temperature of the samples.} \label{table}
\end{table}

\begin{figure}[!ht]
\centerline{\includegraphics[width=8cm]{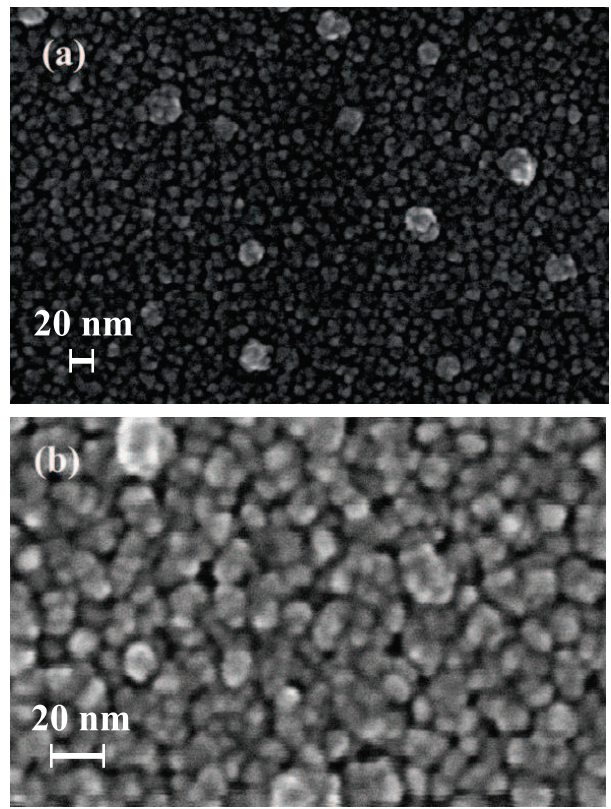}}
\caption{FESEM image of a Nb nanowire network with $d_{Nb}= 8.5$ nm deposited on a PS template with $\oslash= 10$ nm and $\Lambda= 40$ nm recorded at different magnifications, namely around 560 Kx (a) and 1300 Kx (b).} \label{SEM}
\end{figure}

The templates used in this work have a pores average diameter and interpore spacing $\oslash = 5, 10$ nm and $\Lambda  = 10, 40$ nm, respectively. The substrates were obtained by cutting the bare PS substrates in pieces on average 2-3 mm wide and 7-8 mm long. Ultrathin Nb films, with thickness $d_{Nb}= 9$ nm (samples A and B) and $d_{Nb}= 8.5$ nm (sample C), were deposited by UHV sputtering on the PS templates at typical rates of 0.3 nm/s, according to the procedure described in Refs. \cite{Trezza1,Trezza2}. The tendency of the Nb film to occupy only the trenches between the substrate pores generates an array of interconnected wires, namely a network-like structure. Consequently, the wires have a nominally average width $w= \Lambda - \oslash \approx 5$ or $30$ nm which is comparable or even smaller, in an appropriate temperature range, than $\xi$. (Indeed, the Ginzburg-Landau coherence length at zero temperature, estimated from the temperature dependence of the perpendicular upper critical field, $H_{c2\perp}(T)$, is $\xi(0) \approx 10$ nm). These extremely reduced $w$ values implicate that in principle the system may behave as a superconducting wire network in a wide temperature range, and not only in a narrow temperature interval close to the critical temperature, as usually reported \cite{Baturina,Pannetier}. The main samples parameters are summarized in Table \ref{table}. As recently demonstrated \cite{Cirillo,Trezza3}, the transport properties of these systems may show evidence of thermal and quantum fluctuations. Therefore the advantage of this approach is to avoid difficult and expensive nanofabrication steps, allowing the production of high quality quantum objects which, at the same time, are robust and easy to manipulate. 

The morphology of the samples were investigated by field emission scanning electron microscopy (FESEM) analysis, which revealed that the samples are polycrystalline with well shaped grains, whose dimensions are comparable to their thickness. As an example, Fig. \ref{SEM} shows two FESEM images acquired at different magnifications of a sample deliberately fabricated for the morphological analysis, nominally identical to sample C, therefore deposited on a PS substrate with $\oslash= 10$ nm and $\Lambda= 40$ nm. 

The superconducting transport properties were resistively measured on the unstructured samples in a $^{4}$He cryostat using a standard in-line dc four-probe technique. The distance between the silver 
current (voltage) pads is about 5 mm (1 mm). A constant bias current, $I_{b}= 500$ $\mu$A, was applied to the samples. The effective driven current flowing through a single nanowire can be approximately estimated as $I_{b}$ divided by the number of wires contacted below a single contact pad of average diameter of about 1 mm. It results that the bias current flowing through the single nanowire does not exceed $i_{b} \approx 0.5$ nA that corresponds to an average critical current density $j_{b} \approx 1.8$ $\mu$A/m$^2$, about four orders of magnitude smaller than the depairing current density of a single nanowire at $T =$ 0 \cite{Cirillo}. The magnetic field was always applied perpendicularly to the substrate plane. It resulted that the nanowire arrays exhibit well established superconducting properties with a critical temperature, defined as the temperature at which the resistance is reduced of $10\%$ of $R_{N}$ (here $R_{N}$ is the normal state resistance), $T^{10\%}_{c}= 3.00$ K and $3.65$ K for samples A and B, and for sample C, respectively. The resistivity at $T= 10$ K was estimated to be $\rho _{10 K} \approx 50$ $\mu \Omega\cdot$cm. In summary, no signatures of inhomogeneity were evidenced by either FESEM and transport measurements \cite{Cirillo}.  

\begin{figure}[!ht]
\centerline{\includegraphics[width=8cm]{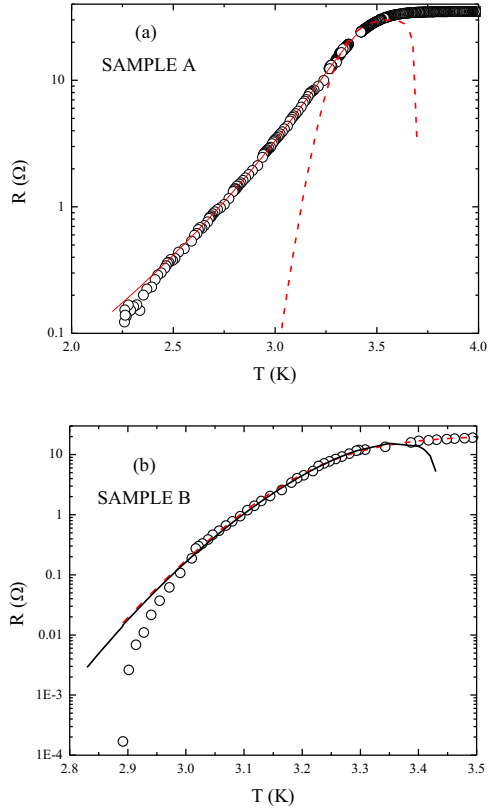}}
\caption{(Color on line) Resistive transitions, $R(T)$, of samples A (a) and B (b). Circles represent the experimental data, the thick solid red line is the theoretical curve obtained according to the complete Eq. \ref{eq:Rtot}, while the thin dashed red line includes only the TAPS term. The thin solid black line represents the best fit to the data obtained with Eq. (\ref{eq:NQUID}).} \label{RT}
\end{figure}

\section{Results and discussion}

The resistive transitions, $R(T)$, of samples A and B are reported in semi-logarithmic scale in Fig. \ref{RT}(a) and Fig. \ref{RT}(b), respectively. The curves do not present any steps revealing fingerprints of inhomogeneity \cite{Bollinger}, but are quite broad, as expected in the presence of fluctuations of the superconducting order parameter. In the case of sample B, the negative curvature present in all the explored temperature range is consistent with a thermal activation scenario, described by the theoretical model proposed for 1D superconductors in the presence of TAPS processes \cite{Zaikin2}:

\begin{equation}
R_{TAPS}(T) \approx R_{Q}\frac{L}{\xi(T)}\frac{T^{*}}{T}\sqrt{\frac{U(T)}{k_{B}T}}\exp\Big[-\frac{U(T)}{k_{B}T}\Big]
\label{eq:TAPS}
\end{equation}

\noindent where $R_{Q}$ is the quantum resistance $R_{Q}=h/4e^{2} \approx 6.45$ k$\Omega$, $T^{*} \approx T_{c}$ is a crossover temperature between TAPS and QPS regime, $L$ is the nanowire length, and $\xi(T)=\xi(0)/\sqrt{1-T/T_{c}}$ and $U(T)=U(0)(1-T/T_{c})^{3/2}$ are the temperature dependent coherence length and phase slip activation energy, respectively. An analogous $R(T)$ dependence (not shown here) is obtained for sample C. On the other hand, sample A, due to the extremely reduced values of its effective wire diameter $\sigma= 6.7$ nm ($\sigma= \sqrt{d_{Nb} \cdot w}$) compared to samples B and C, shows a pronounced resistance tail, characteristic of quantum fluctuations processes. The latter can be described by the relation \cite{Zaikin1,Zaikin3,Bae,Lehtinen}:

\begin{equation}
R_{QPS}(T) \approx \alpha \frac{R_{Q}^{2}}{R_{N}}\frac{L^{2}}{\xi^{2}(0)}\exp\Big[-\alpha \frac{R_{Q}}{R_{N}}\frac{L}{\xi(T)}\Big] \label{eq:QPS}
\end{equation}

\noindent where $\alpha$ is a fitting parameter of the order of the unity. The total resistance of the samples can be expressed as \cite{Lau}:

\begin{equation}
R(T) = [R_{N}^{-1} + (R_{TAPS}+R_{QPS})^{-1}]^{-1}  \label{eq:Rtot}
\end{equation}

\noindent where the QPS term will be present only in the analysis of sample A. 
Indeed, the results of the theoretical interpretation of the $R(T)$ transitions according to the complete Eq. \ref{eq:Rtot} for sample A is displayed by the solid red line in Fig. \ref{RT}(a). The excellent accordance between the theoretical curve and the experimental data confirms the hypothesis of the presence of QPS. This conclusion is reinforced by comparing the points with the red dashed line, obtained disregarding the QPS term in Eq. \ref{eq:Rtot}. On the contrary, the best fit curve that reproduces the data of sample B includes only the TAPS contribution. The latter is represented by the red dashed curve in Fig. \ref{RT}(b) and it nicely follows the data down to $T \approx 3$ K. Indeed, the negative curvature of the $lnR(T)$ data is not consistent with a QPS scenario, therefore the tentatives (not shown here) to fit the resistive tail taking into account quantum fluctuations were not satisfactory. The poor agreement between theory and experiment, compared with the results reported in Ref. \cite{Cirillo,Trezza3}, can be possibly attributed to the fact that the samples analyzed here are not structured. Indeed structuring enables to obtain better controlled arrays with a reduced number of interconnected wires \cite{Cirillo,Trezza3} and, consequently, a narrower distribution of widths and activation energies. It is worth reminding, in fact, that Eq. \ref{eq:Rtot} is strictly valid for a single nanowire. However, the value of the activation energy extracted here, $U(0)= 3.9$ meV, does not differ much from the value estimated in Ref. \cite{Cirillo}, while it is quite different from the one derived from the theoretical expression $U^{th}(0) \approx 0.83 (\sigma^2 /\xi(0)) (R_{Q}/\rho _{10K}) k_{B}T_{c} \approx 75$ meV again valid for an individual nanowire  \cite{Bae}. This discrepancy, as the similar one reported in Ref. \cite{Luo}, is not surprising since in both cases the behavior of superconducting wire networks is described by models derived for single wires.

\begin{figure}[!ht]
\centerline{\includegraphics[width=8cm]{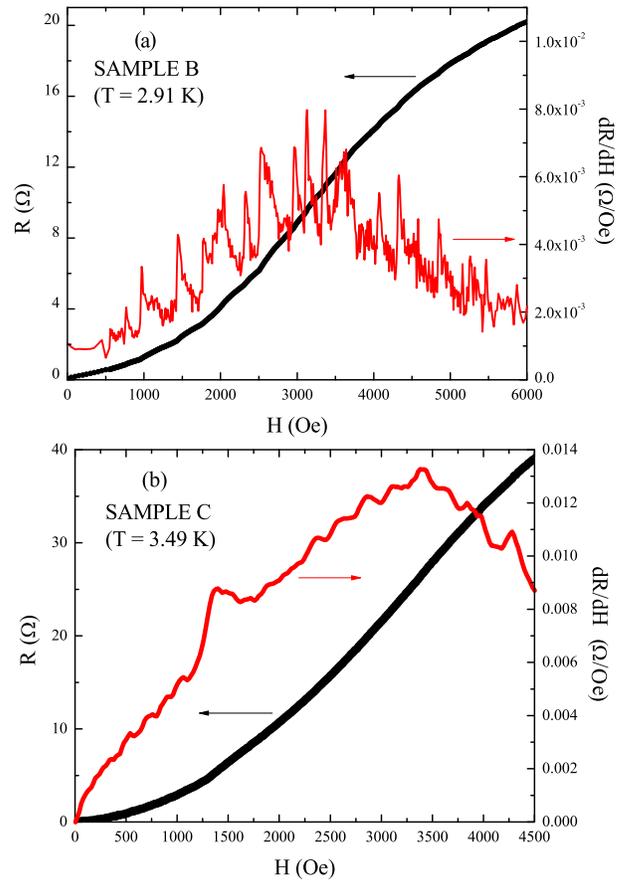}}
\caption{(Color online). (a)-(b): On the left (right) scale the magnetoresistance transitions, $R(H)$, (first magnetoresistance derivative, $dR/dH (H)$) are shown for samples B (a) and C (b).} \label{derivative}
\end{figure}

Fig. \ref{derivative} shows the central result of this paper, namely the oscillatory field dependence of the $R(H)$ transitions. In panel $(a)$ this result is presented for sample B at $T= 2.91$ K. The effect is disclosed superimposing on the right scale the first $R(H)$ derivative as a function of the field, $dR/dH$, showing that the oscillations are present in all the transition range, up to $R_N$. Similar magnetoresistance oscillations were observed also for samples A and C. As an example, in panel $(b)$ the $R(H)$ together with its derivative are reported at $T= 3.49$ K for sample C. 

\begin{figure}[!ht]
\centerline{\includegraphics[width=8cm]{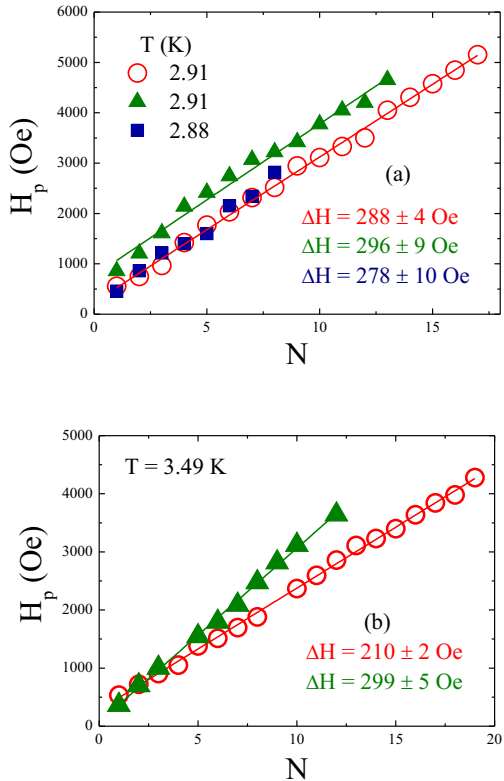}}
\caption{(Color online). Peaks positions, $H_p$, as a function of the index number, N, for samples B (a) and C (b). (a) Open red (closed green and blue) symbols refer to peaks positions extracted from the $dR/dH (H)$ curve of Fig. \ref{derivative}(a) ($\Delta R(H)$ curves in Fig. \ref{subtraction}(b)). The slopes of the best fit lines are reported in the legend. (b) Peaks positions extracted from the $dR/dH (H)$ curve of Fig. \ref{derivative}(b), (open symbols) and from the $\Delta R(H)$ curve in Fig. \ref{subtraction}(c) (closed symbols). The slopes of the best fit lines are reported in the legend.} \label{peaks}
\end{figure}

While the shape of the resistive transitions indicates the strong influence of fluctuations, the MR oscillations are fingerprints of a coherent state and demonstrate the multiple connectedness of the samples.
Due to the network-like structure of these samples, the $R(H)$ experimental data could remind the Little-Parks effect. However, for all the samples, the values of the oscillations periods are not compatible with this interpretation. In Fig. \ref{peaks}(a) by open symbols is shown the indexation of the field at which peaks, $H_p$, are present in the $dR/dH$ curve for sample B in Fig. \ref{derivative}(a). The slope of the resulting best fit line represents the oscillations period, which in this case is $\Delta H = 288$ Oe. According to the relation $\Delta H = \Phi_{0}/S$ this value would correspond to an elemental area of the network $S= 7 \times 10^{-14}$ m$^{2}$, so that the pore diameter in the framework of the LP scenario is expected to be $\oslash^{LP} \approx 2 (S/\pi )^{1/2}\approx  300$ nm. However, this value is one order of magnitude larger than the pores of the present templates, $\oslash$. Similar considerations are also valid in the case of sample C for which, using the same approach, an oscillation period of $\Delta H = 210$ Oe was estimated (open symbols in Fig. \ref{peaks}(b)). These numbers are incompatible also if one tries to identify the MR structures as fractional fluxoid quantization effects per unit cell as reported for a square or a triangular network \cite{Pannetier,Nori}. The possible distortion of the supercurrents in the multiconnected samples was also considered as possible explanation of this unexpected result. According to Ref. \cite{Zhang2} the effective current path could have an elliptical shape around an effective area $S_{eff}=\pi (\oslash + w/2)(\oslash + \xi(t)/2)$. At the reduced temperatures of Figs. \ref{derivative}(a) and (b) this would correspond to the oscillation periods $\Delta H \approx 6800$ Oe and $\Delta H \approx 7500$ Oe for samples B and C, respectively, again in disagreement with the observed ones. The same discrepancies occur if one tries to interpret these oscillations as due to intra-grains screening currents \cite{Herzog,Patel2} since, in this thickness regime, typical grains dimensions of all the analyzed samples are expected to be comparable to the film thickness \cite{Cirillo}. In summary, it is not possible to easily identify this expected coherent area of the superconducting sample with any characteristic length scale of the structure. It is also worth reminding that, despite the similarity with the systems investigated in Ref. \cite{Luo}, the anomalies in the MR data reported there and interpreted as a fingerprint of the LP effect, appear different from the ones object of this study. 

\begin{figure}[!ht]
\centerline{\includegraphics[width=8cm]{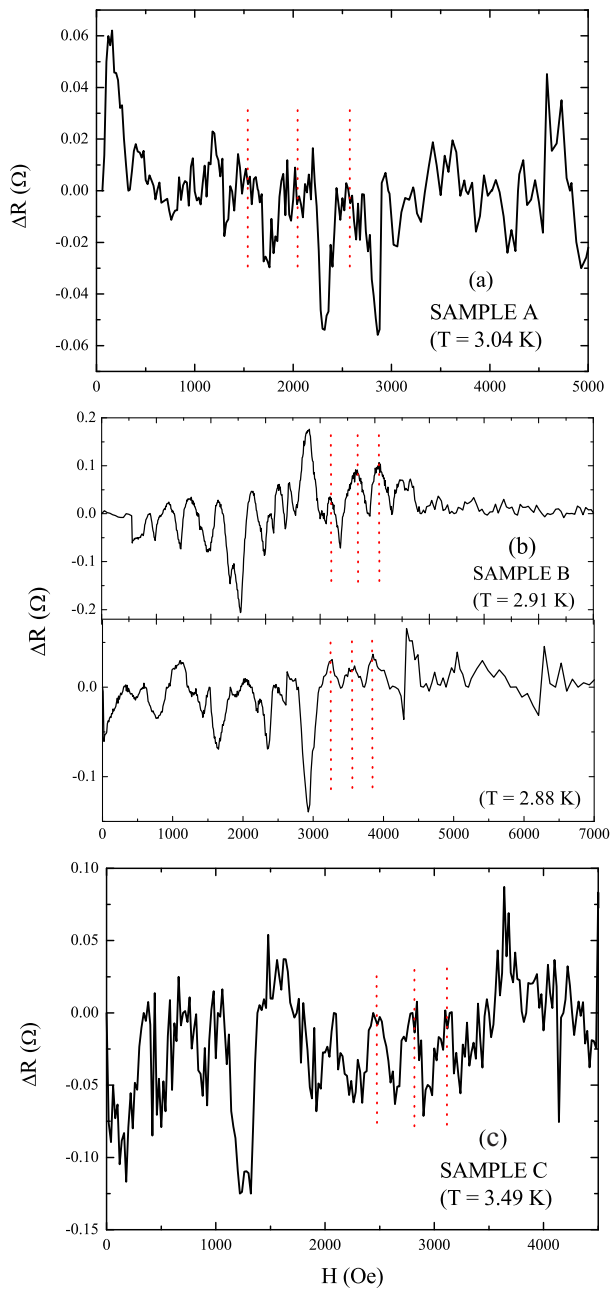}}
\caption{(Color online). $\Delta R=R(H)-R(H)^{int}$ data (see the text for the definition) for all the investigated samples at some selected different temperatures. Red dotted lines indicates a selection of peak positions, $H_p$.} \label{subtraction}
\end{figure}

 Similarly to Figs. 2 and 3 of Ref. \cite{Johansson}, in order to better appreciate the oscillatory $R(H)$ dependence initially recognized in the $dR/dH$ curves, a smooth background was subtracted from the original $R(H)$ curves. The adopted procedure is described in the following. First the envelope of the $R(H)$ curves, $R(H)^{int}$, was obtained by a linear interpolation with roughly the same point density for all the samples. Then an Origin tool \cite{Origin} is used to subtract the $R(H)^{int}$ curve from the $R(H)$ one. The results of the data manipulation, $\Delta R=R(H)-R(H)^{int}$, are displayed in Fig. \ref{subtraction} for all the investigated samples at some selected different temperatures. It is evident that the oscillatory behavior has a periodic or possibly a multiple periodic character, due to the presence of many possible current loops \cite{Hansma}. In the following part of the paper the analysis will focus on the largest periodicities corresponding to the smallest areas, as highlighted, for the sake of the clarity, in all the panels by red dotted lines that indicate a selection of these peaks positions, $H_p$. These periodic features are well detectable up to $H \approx 4000$ Oe for all the samples. At the analyzed temperatures the oscillation periods were estimated from the linear fitting of these peak positions, resulting in $\Delta H \approx 500$ Oe for sample A, and $\Delta H \approx 300$ Oe for samples B (Fig. \ref{peaks}(a)) and C (Fig. \ref{peaks}(b)), respectively. Moreover, in the case of sample B (Fig. \ref{subtraction}(b)) one can also observe that, despite the small temperature range investigated, the sharpness of the oscillations seems to reduce with the decrease of  $T$, while the opposite result was reported in the case of granular samples \cite{Herzog}. On the contrary, both the position of the peaks and the values of the periods are not affected by $T$. This last result is displayed in Fig. \ref{peaks}(a), where the indexation of the peaks, obtained from the $\Delta R (H)$ curves, is reported by solid symbols for both $T= 2.91$ K and $T= 2.88$ K (triangles and squares, respectively). From the slope of the linear fitting curves to the data it follows that the periods, evaluated from the $dR/dH$ and the $\Delta R$ curves, are in perfect agreement within the experimental error (see legend in the Figure). A discrepancy is, instead, present in the case of sample C, since from the analysis of the $dR/dH$ data (open symbols in Fig. \ref{peaks}(b)) a value of $\Delta H= 210$ Oe was extracted, while from the $\Delta R (H)$ curve it follows $\Delta H= 299$ Oe (closed triangles in Fig. \ref{peaks}(b)). However, it is reasonable to assume that the presence of the pronounced anomaly around $H \approx 1500$ Oe, related to a fractional matching effect \cite{Trezza2}, may mask the presence of some peaks, affecting the analysis. 

Due to the multiple connectedness of the samples, this work presents several analogies with the ones of Refs. \cite{Hansma,Saxena,Chernov} that deal with arrays of superconducting particles, where, indeed, multiply periodic dependence of the critical current \cite{Hansma} or regular oscillations of the induced voltage \cite{Saxena,Chernov} on the magnetic field were observed. These kind of systems were successfully modeled in Ref. \cite{Hansma} as an array of Josephson junctions and behaves similarly to a conventional double junction interferometer. In this case the MR oscillations could have origin in the formation of $i$ different conduction paths enclosing an area $S_i$, therefore resulting in different periodicities $\Delta H_i = \Phi_{0}/S_i$. Indeed, particularly Fig. \ref{derivative}(a) discloses signatures of multiple periodicities, so that, drawing a parallel between these systems and the nanoparticles arrays \cite{Hansma,Saxena,Chernov}, these oscillations periods should correspond to the smallest current path with an area of the order of $S_{min}=4-9 \times 10^{-14}$ m$^{2}$ and a radius of the order of $3-4$ times $\Lambda $.

In analogy, a more advanced device based on two short NWs in a Dayem bridge configuration was proposed in Refs. \cite{BezryadinScience,Pekker} and was modeled as a pair of current-biased Josephson junctions. In this case, the system consists of a couple of parallel superconducting nanowires of length $b$ at a distance $2a$ from each other, suspended over two mesoscopic superconducting leads of width $2l$ (see Fig. 2 of Ref. \cite{Pekker}), obtained by a highly sophisticated fabrication method \cite{BezrNanotech}. Since $\xi < 2l < \lambda_{\perp}$ (where $\lambda_{\perp}=\lambda^{2}/d$ is the perpendicular penetration depth), the external field penetrates the leads, producing phase gradients which, in turn, induce supercurrents in the wires. The wires behave as one-dimensional objects governed by TAPS processes, but due to the geometry of the systems, constraints are imposed on the phase of the order parameter in the device. Therefore, according to the model proposed by the authors, it results that the magnetoresistance period of the device is determined not only by the flux concatenated to the area $2ab$, but also by the effective area $4al$:

\begin{equation}
\Delta H = \Big[\Big(\frac{\Phi_{0}}{4al}\Big)^{-1}+\Big(\frac{\Phi_{0}}{2ab}\Big)^{-1}\Big]^{-1}
\label{eq:period}
\end{equation}

The first term in Eq. \ref{eq:period} accounts for the short oscillation period. 
The authors named this device NQUID, since it acts as a (Superconducting) Nanowire Quantum Interference Device. 
In the limit of short NWs ($b \ll l$), the resistance of the two-wires device is approximated as \cite{Pekker}: 

\begin{eqnarray}
R_{NQUID} & = & 2R_{N} \frac{\sqrt{(1-x^{2})}}{x} \times \nonumber \\ &
\times & \exp[-\gamma(\sqrt{(1-x^{2})}+x sin^{-1}x)] \times \nonumber \\ &
\times & sinh(\frac{\pi \gamma x}{2})
\label{eq:NQUID}
\end{eqnarray}
\noindent where $x=I/I_{c}$, $\gamma=\hbar I_{c}/e k_B T$ and with $I_{c}=\sqrt{(I_{c1}+I_{c2})^{2} cos^{2}\delta + (I_{c1}-I_{c2})^{2} sin^{2}\delta}$ and $\delta=-4 \pi alH/\Phi_0$. Here $I$ is the bias current, and $I_{c1,2}$ are the critical currents of the wires. Eq. (\ref{eq:NQUID}) describes both the temperature and the field dependence of the two-wires device and in principle can be extended to the more general case of a multi-wire device \cite{Pekker}.

A parallel can be drawn between the system under study and the device described in Refs. \cite{BezryadinScience,Pekker}. In particular it appears straightforward that it possesses one of the two key ingredients necessary to the operation of the NQUID, namely superconducting NWs showing dissipative fluctuations. As it results from the theoretical analysis of the curves in Fig. \ref{RT}, in these samples both thermal and quantum fluctuations may be important. On the other hand, the operation of the NQUID also requires the presence of mesoscopic leads. Here it is evident that this high-field periodicity corresponds to coherence over large areas. As it emerged from the FESEM analysis, due to the peculiar growth technique of the network, the template coverage may be not uniform and locally some small continuous area may appear, in particular when $d_{Nb} \approx \oslash$.  Indeed it is also worth reminding that, from a previous work \cite{Cirillo}, it resulted that great care must be paid to the choice of the film thickness in order to assure both wires continuity and network definition. From these considerations it follows that in the transition regime (before the resistance goes to zero), where the coherence length is large, grains may be effectively coupled over significant lengths, and may act as a sort of mesoscopic superconducting lead. Also in this case the analogy with the works on particles arrays \cite{Saxena,Chernov} is evident, since in these systems the agglomerates of particles with dimensions $D_{i} \approx S_{i}^{1/2}$ are responsible of the MR oscillation. Interestingly, also the temperature interval where the effect is present is limited to the transition regime in this network as well as in Ref. \cite{Saxena}. Indeed, here the oscillations are present over the range of temperatures in which $R$ approaches zero and vanish as the temperature is further lowered. It is worth noticing that the good coupling among the grains was already demonstrated in Ref.~\cite{Cirillo}, where it was also excluded that weak links are responsible of the conduction. Moreover, the extreme reduced thickness of the Nb films assures that the penetration depth is larger than $d_{Nb}$ in all the investigated $T$ range, and therefore the sample is always transparent to the external magnetic field. Indeed, even if one considers the London penetration depth of bulk Nb: $\lambda_{L} (0) \approx 40$ nm \cite{Cyrot}, it results that at $T = 0$, $\lambda_{\perp} (0) =\lambda_{L}^{2} (0) /d \approx 170$ nm, a value which is by far larger than the characteristic dimensions of the considered networks.

In order to try to deepen the possible analogy between these samples and the proposed NQUID, as in the case of sample B, where indeed TAPS dominate the transport at low temperature, the experimental data were reproduced also according to Eq.~\ref{eq:NQUID} adapted to the multiwire case proposed in the Appendix of Ref. \cite{Pekker}. This approach is valid since the nanowires are shorter than the superconducting coherence length in the analyzed temperature regime. By imposing $H= 0$, it follows that $\delta= 0$ and the expression for $I_{c}$ reduces to $I_{c}= \Sigma_i I_{ci}$, namely it represents the sum of the critical currents of all the single NWs in the samples. The fitting procedure was performed treating $I_{c}$, $I$, and $T_c$ as fitting parameters. Considering $I_{c}$ as a whole is consistent with the extension to the multiwire device obtained assuming that all the wires have identical critical currents. Since the assumption of the theory reported in Refs. \cite{BezryadinScience,Pekker} is that the wires resistance is due to thermally activated phase slips, it is not surprising that the best fitting curve obtained according to Eq. (\ref{eq:NQUID}) does not differ much from the previous approach. The solid black curve reported in Fig. \ref{RT}(b) represent the best fit obtained with the following values of the parameters: $I_{c}= 7.6$ $\mu $A, $I= 0.1$ $\mu $A, and $T_c= 3.67$ K. The values extracted from the fitting procedure deserve a comment. First, the value of $T_c$ is close to the onset of the resistive transition, but, while the value of $I_{c}$ is of the same order of magnitude measured for the analogous systems investigated in Refs. \cite{Cirillo,Trezza3}, the one of $I$ is considerably higher than the figure obtained from the crude approximation used to estimate the effective bias current $i_{b}$ reported above.
It is also worth mentioning that the same fitting procedure was not performed in the case of sample A, since within the model of Refs. \cite{BezryadinScience,Pekker} it is assumed that the resistance fluctuations are due to TAPS only. However, there is no reason to believe that the same device would not operate in the case of two NWs exhibiting QPS.  

With the aim of establishing a comparison between the nanowire networks and the above mentioned two-wires device \cite{BezryadinScience,Pekker}, it is possible to estimate the dimensions of the area over which the superconducting Nb grains should be coupled from Eq. \ref{eq:period}. As a first approximation, identifying $a$ with $\Lambda $ it results $l \approx 1$ $\mu$m for sample A and $l \approx 410$ nm for both samples B and C. At a first glance these values may appear far too high. However, one has to keep in mind that the $R(H)$ measurements were performed $in$ the transition, where the values of the superconducting coherence length are extremely large. In particular, in the case of sample A, the MR was acquired at a temperature higher than the chosen definition of $T_c$. It is also important to point out that here the effect is present in an extremely narrow temperature range and disappears with the decreasing of $T$. This temperature dependence deserves a comment, since it can be useful to rule out the interpretation of the observed MR effect in the framework of different scenarios. Indeed, MR oscillations caused by intra-grains circulating currents \cite{Herzog,Patel2}, non-uniform sample thickness \cite{Johansson,Wang1,Wang2}, or by randomly distributed pinning centers \cite{Mills} are usually more and more pronounced as the temperature is lowered. On the other hand, similarly to the present case, both the two-wire device of Ref. \cite{Pekker} and the arrays of Ref.  \cite{Saxena} show peculiar MR oscillations in the temperature range of the resistive transition, which, in presence of phase slips, can in principle be rather broad.  
However, discrepancies between the present superconducting networks and the other quoted systems \cite{BezryadinScience,Pekker,Hansma,Saxena,Chernov} are noticeable in terms of  values of the MR oscillations periods. While here $\Delta H \approx 300$ Oe, in the two-wires device \cite{BezryadinScience,Pekker} the largest observed period is lower than 10 Oe. On the other hand, for the same system the so-called high-field regime \cite{BezryadinScience}, corresponding to the regime when vortices enter the leads, was explored. In this regime both the MR and the switching current of the device present an oscillatory behavior with a field-dependent period which in both cases agrees with the estimation  $\Delta H_{large}= \Phi_{0}/ 2a(d+b)$, where $d$ is the distance between the vortices. The MR oscillations observed here cannot be interpreted in this framework and the reason could be twofold. First the experimental values of $\Delta H$ do not agree with $\Delta H_{large}$ for any reasonable values of $d$, and second, while the inter-vortex distance varies considerably in the wide analyzed field range, $\Delta H$ does not depend on $H$. Due to this last discrepancy and to the main concern related to the absence of the leads, a quantitative comparison of the MR data with Eq. (\ref{eq:NQUID}) was not accomplished. However, in order to deepen the comprehension of the MR oscillations, a properly designed experiment in which a smaller portion of the network is biased via two continuous superconducting mesoscopic banks would be interesting.

\section{Conclusions}

In conclusion, MR oscillations were observed in superconducting NWs networks fabricated by a low-cost alternative approach based on the deposition of ultrathin high quality Nb films on self-assembled PS templates. The interest on the proposed samples is not only of fundamental, but also of applicative nature, since their operation mimics the one of a SQUID device. Despite the exact mechanism of the current distribution are not completely clarified, it is now clear that these systems offer the unique opportunity of accessing the low dimensional regime using a broadly accessible fabrication technique. Indeed, in the next future two experiments are planned. First, electron beam lithography could be used to fabricate much narrower bridges and/or more dedicated structures, obtaining a better control of the MR oscillation period. Second, it would be interesting to study the response of the sample to a RF radiation, both as a function of the irradiation power and of the applied magnetic field. These steps are compulsory to determine possible application on a longer term of the NWs networks presented here. In particular, they could be appealing for the realization of superconducting quantum interference devices with a proper spatial resolution capable of the detection of small spin systems, since the spin sensitivity scales with the SQUID loop radius \cite{Granata}. 

\section{Acknowledgements}

The authors gratefully acknowledge A. Vecchione for the FESEM images and G. Carapella for both the careful reading of the manuscript and the fruitful discussions.


\vspace{0.5in}



\begin{references}

\bibitem{Tinkham} M. Tinkham, Journal of Superconductivity: Incorporating Novel Magnetism {\bf13}, 801 (2000).

\bibitem{nanopart} S. Bose, A.M. Garcia-Garcia, M.M. Ugeda, J.D. Urbina, C.H. Michaelis, I. Brihuega, and K. Kern, Nature Materials {\bf9}, 550 (2010).

\bibitem{AruPhysRep} K.Y. Arutyunov, D.S. Golubev, and A.D. Zaikin, Physics Reports {\bf464}, 1 (2008).

\bibitem{BezBook} A. Bezryadin, Superconductivity in Nanowires: Fabrication and Quantum Transport (Wiley-VCH, Germany, 2013).

\bibitem{Rodrigo} J.G. Rodrigo, V. Crespo, H. Suderow, S. Vieira, and F. Guinea, Phys. Rev. Lett. {\bf109}, 237003 (2012).

\bibitem{Hopkins} D.S. Hopkins, D. Pekker, T.-C. Wei, P.M. Goldbart, and A. Bezryadin, Phys. Rev. B {\bf76}, 220506(R) (2007).

\bibitem{Delacour} C. Delacour, B. Pannetier, J.-C. Villegier, and V. Bouchiat, Nano Lett. {\bf12}, 3501 (2012).

\bibitem{Murphy} A. Murphy, P. Weinberg, T. Aref, U.C. Coskun, V. Vakaryuk, A. Levchenko, and A. Bezryadin, Phys. Rev. Lett. {\bf110}, 247001 (2013).

\bibitem{Weides} M. Weides and H. Rotzinger, Advanced superconducting circuits and devices in Handbook of Applied Superconductivity (Wiley-VCH, Germany, 2014).

\bibitem{Granata} C. Granata, E. Esposito, A. Vettoliere, L. Petti, and M. Russo,  Nanotechnology {\bf19}, 275501 (2008).

\bibitem{BezryadinScience} D.S. Hopkins, D. Pekker, P.M. Goldbart, and A. Bezryadin, Science {\bf308}, 1762 (2005).

\bibitem{Pekker} D. Pekker, A. Bezryadin, D.S. Hopkins, and P.M. Goldbart, Phys. Rev. B {\bf72}, 104517 (2005).

\bibitem{Hansma} P.K. Hansma and J.R. Kirtley, J. Appl. Phys. {\bf45}, 4016 (1974).

\bibitem{Nazarov} J.E. Mooj and Y.V. Nazarov, Nat. Phys. {\bf2}, 169 (2006).

\bibitem{Astafiev} O.V. Astafiev, L.B. Ioffe, S. Kafanov, Y.A. Pashkin, K.Y. Arutyunov, D. Shahar, O. Cohen, and J.S. Tsai, Nature {\bf484}, 355 (2012).

\bibitem{Mooij} J.E. Mooij and C. Harmans, New J. Phys. {\bf7}, 219 (2005).

\bibitem{Belzig} G. Rastelli, M. Vanevic, and W. Belzig, arXiv:1403.4565v1 (2014).

\bibitem{Hongisto} T.T. Hongisto and A.B. Zorin, Phys. Rev. Lett. {\bf108},  097001 (2012). 

\bibitem{Webster} C.H. Webster, J.C. Fenton, T.T. Hongisto, S.P. Giblin, A.B. Zorin, and P.A. Warburton, Phys. Rev. B {\bf87},  144510 (2013). 

\bibitem{AruNanotech} M. Zgirski, K.-P. Riikonen, V. Tuboltsev, P. Jalkanen, T.T. Hongisto, and K.Yu Arutyunov, Nanotechnology {\bf19}, 055301 (2008).

\bibitem{Tettamanzi} G.C. Tettamanzi, C.I. Pakes, A. Potenza, S. Rubanov, C.H. Marrows, and S. Prawer, Nanotechnology {\bf20}, 465302 (2009).

\bibitem{BezrNanotech} M. Remeika and A. Bezryadin, Nanotechnology {\bf16}, 1172 (2005).

\bibitem{Tian} M. Tian, J. Wang, J.S. Kurtz, Y. Liu, M.H.W. Chan, T.S. Mayer, and T.E. Mallouk, Phys. Rev. B {\bf71}, 104521 (2005).

\bibitem{Luo} Q. Luo, X.Q. Zeng, M.E. Miszczak, Z.L. Xiao, J. Pearson, T. Xu, and W.K. Kwok, Phys. Rev. B {\bf85}, 174513 (2012).

\bibitem{BezAPL} A. Rogachev and A. Bezryadin, Appl. Phys. Lett. {\bf83}, 512 (2003).

\bibitem{Giordano} N. Giordano, Phys. Rev. B {\bf41}, 6350 (1990).

\bibitem{BezryadinNat} A. Bezryadin, C.N. Lau, and M. Tinkham, Nature {\bf404}, 971 (2000).

\bibitem{Lau} C.N. Lau, N. Markovic, M. Bockrath, A. Bezryadin, and M. Tinkham, Phys. Rev. Lett. {\bf87}, 217003 (2001).

\bibitem{Sochnikov} I. Sochnikov, A. Shaulov, Y. Yeshurun, G. Logvenov, and I. Bo\v{z}ovi\'{c}, Nature Nanotech. {\bf5}, 516 (2010).

\bibitem{Baturina} T.I. Baturina, V.M. Vinokur, A.Yu. Mironov, N.M. Chtchelkatchev, D.A. Nasimov, and A.V. Latyshev, Europhys. Lett. {\bf93}, 47002 (2011).

\bibitem{Zhang1} W.J. Zhang, S.K. He, H. Xiao, G.M. Xue, Z.C. Wen, X.F. Han, S.P. Zhao, C.Z. Gu, and X.G. Qiu, Physica C {\bf480}, 126 (2012).

\bibitem{Zhang2} W.J. Zhang, S.K. He, H.F. Liu, G.M. Xue, H. Xiao, B.H. Li, Z.C. Wen, X.F. Han, S.P. Zhao, C.Z. Gu, X.G. Qiu, and V.V. Moshchalkov, Europhys. Lett. {\bf99}, 37006 (2012).

\bibitem{Herzog} A.V. Herzog, P. Xiong, and R.C. Dynes, Phys. Rev. B {\bf58}, 14199 (1998).

\bibitem{Johansson} A. Johansson, G. Sambandamurthy, D. Shahar, N. Jacobson, and R. Tenne, Phys. Rev. Lett. {\bf95}, 116805 (2005).

\bibitem{Patel1} U. Patel, S. Avci, Z.L. Xiao, J. Hua, S.H. Yu, Y.Ito, R. Divan, L.E. Ocola, C. Zheng, J. Hiller, U. Welp, D.J. Miller, and W. K. Kwok, Appl. Phys. Lett. {\bf91}, 162508 (2007).

\bibitem{Wang1} J. Wang, X.-C. Ma, L. Lu, A.-Z. Jin, C.-Z. Gu, X.C. Xie, J.-F. Jia, X. Chen, and Q.-K. Xue, Appl. Phys. Lett. {\bf92}, 233119 (2008).

\bibitem{Wang2} J. Wang, X. Ma, S. Ji, Y. Qi, Y. Fu,A. Jin, L. Lu, C. Gu, X.C. Xie, J.-F. Jia, X. Chen, and Q.-K. Xue, Nano Res. {\bf2}, 671 (2009).

\bibitem{Patel2} U. Patel, Z. L. Xiao, A. Gurevich, S. Avci, J. Hua, R. Divan, U. Welp, and W. K. Kwok, Phys. Rev. B {\bf80}, 012504 (2009).

\bibitem{Lehtinen1} J.S. Lehtinen and K. Yu Arutyunov,  Supercond. Sci. Technol. {\bf25}, 124007 (2012).

\bibitem{Mills} S.A. Mills, N.E. Staley, J.J. Wisser, C. Shen, Z. Xu, and Y. Liu, Appl. Phys. Lett. {\bf104}, 052604 (2014).

\bibitem{Peeters} G.R. Berdiyorov, M.V. Milo\v{s}evic, M.L. Latimer, Z.L. Xiao, W.K. Kwok, and F.M. Peeters, Phys. Rev. Lett. {\bf109}, 057004 (2012).

\bibitem{Pannetier} B. Pannetier, J. Chaussy, R. Rammal, and J.C. Villegier, Phys. Rev. Lett. {\bf53}, 1845 (1984).

\bibitem{Kato} M. Kato and O. Sato, Supercond. Sci. Technol. {\bf26}, 033001 (2013) and references therein.

\bibitem{Little} R.D. Parks and W.A. Little, Phys. Rev. A {\bf133}, 97 (1964).

\bibitem{Pavesi} O. Bisi, S. Ossicini, and L. Pavesi, Surf. Sci. Rep. {\bf38}, 1 (2000).

\bibitem{Ge} M. Ge, X. Fang, J. Rong, and C. Zhou, Nanotechnology {\bf24}, 422001 (2013).

\bibitem{PSL} S.L. Prischepa, A.L. Dolgiy, H.V. Bandarenka, V.P. Bondarenko, K.I. Yanushkevich, V.G. Bayev, A.A. Maximenko, Yu. A. Fedotova, A. Zarzycki, and Y. Zabila, in  Nanowires: Synthesis, Electrical Properties and Uses in Biological Systems (Luke J. Wilson, Nova Sci., New York, 2014).

\bibitem{Dolgiy12} A. Dolgiy, S.V. Redko, H. Bandarenka, S.L. Prischepa, K. Yanushkevich, P. Nenzi, M. Balucani, and V. Bondarenko, J. Electrochem. Soc. \textbf{159}, D623 (2012).

\bibitem{Panarin10} Yu. Panarin, S.N. Terekhov, K.I. Kholostov, and V.P. Bondarenko, Appl. Surf. Sci. \textbf{256}, 6969 (2010).

\bibitem{Trezza1} M. Trezza, S.L. Prischepa, C. Cirillo, R. Fittipaldi, M. Sarno, D. Sannino, P. Ciambelli, M.B.S. Hesselberth, S.K. Lazarouk, A.V. Dolbik, V.E.
Borisenko, and C. Attanasio, J. Appl. Phys. {\bf104}, 083917 (2008).

\bibitem{Trezza2} M. Trezza, C. Cirillo, S.L. Prischepa, and C. Attanasio, Europhys. Lett. {\bf88}, 57006 (2009).

\bibitem{Cirillo} C. Cirillo, M. Trezza, F. Chiarella, A. Vecchione, V.P. Bondarenko, S.L. Prischepa, and C. Attanasio, Appl. Phys. Lett. {\bf101}, 172601 (2012).

\bibitem{Trezza3} M. Trezza, C. Cirillo, P. Sabatino, G. Carapella, S.L. Prischepa, and C. Attanasio, Appl. Phys. Lett. {\bf103}, 252601 (2013).

\bibitem{Bollinger} A.T. Bollinger, A. Rogachev, M. Remeika, and A. Bezryadin, Phys. Rev. B {\bf69}, 180503(R) (2004).

\bibitem{Zaikin2} D.S. Golubev and A.D. Zaikin, Phys. Rev. B. {\bf78}, 144502 (2008).

\bibitem{Zaikin1} A.D. Zaikin, D.S. Golubev, A. van Otterlo, and G.T. Zim\'{a}nyi, Phys. Rev. Lett. {\bf78}, 1552 (1997).

\bibitem{Zaikin3} D.S. Golubev and A.D. Zaikin, Phys. Rev. B {\bf64}, 014504 (2001).

\bibitem{Bae} M.-H. Bae, R.C. Dinsmore III, T. Aref, M. Brenner, and A. Bezryadin, Nano Lett. {\bf9}, 1889 (2009).

\bibitem{Lehtinen} J.S. Lehtinen, T. Sajavaara, K.Yu. Arutyunov, M.Yu. Presnjakov, and A.L. Vasiliev, Phys. Rev. B. {\bf85}, 094508 (2012).

\bibitem{Nori} Q. Niu and F. Nori, Phys. Rev. B. {\bf39}, 2134 (1989).

\bibitem{Origin} www.originlab.com/doc/X-Function/ref/subtract\_ref

\bibitem{Saxena} A.M. Saxena, J.E. Crow, and M. Strongin, Solid State Commun. {\bf14}, 799 (1974).

\bibitem{Chernov} A.V. Gabrel'yan, Y.G. Morozov, and E.A. Chernov, Solid State Commun. {\bf65}, 889 (1988).

\bibitem{Cyrot} M. Cyrot and D. Pavuna, Introduction to Superconductivity and High-T$_c$ Materials (World Scientific, Singapore, 1992).



\end{references}
\end{document}